\documentstyle[amssymb]{amsart}

\def\fun#1#2{\lower3.6pt\vbox{\baselineskip0pt\lineskip.9pt
  \ialign{$\mathsurround=0pt#1\hfil##\hfil$\crcr#2\crcr\sim\crcr}}}
\newskip\humongous \humongous=0pt plus 1000pt minus 1000pt

\newif\ifdtup

\def\lqqp{\lq \lq \ }
\def\lqq{\lq \lq }
\relax

\newtheorem{THEOREM}{Theorem}[section]
\newtheorem{LEMMA}{Lemma}[section]

\theoremstyle{definition}
\newtheorem{definition}{Definition}[section]

\newcommand{\be}{\begin{equation}}
\newcommand{\ee}{\end{equation}}

\addtocounter{equation}{-1}
\begin{document}

\rightline{Preprint IFUNAM FT 94-57}
\rightline{July 1994}
\vskip 1.truecm

\title[Lie-algebras and quasi-exactly \dots]
{Quasi-exactly-solvable differential equations}

\author{ Alexander Turbiner}
\address
{Institute for Theoretical and Experimental Physics,
Moscow 117259, Russia}
\email{ TURBINER@@CERNVM or TURBINER@@VXCERN.CERN.CH}
\curraddr{Instituto de F\'isica, Universidad Nacional Aut\'onoma
de M\'exico, Apartado Postal 20-364, 01000 M\'exico D.F., MEXICO}
\thanks{Supported in Part by a CAST grant of the US National Academy of
Sciences and a research grant of CONACyT, Mexico}
\thanks{This paper will be published in CRC Handbook of Lie Group Analysis
of Differential Equations, Vol. 3 (New Trends) , Chapter 12, CRC Press,
N. Ibragimov (ed.)}

\date{}
\maketitle
\begin{abstract}
A general classification of linear differential
 and finite-difference operators possessing a finite-dimensional
invariant subspace with a polynomial basis is given. The main result
is that any operator with the above property must have a representation
as a polynomial element of the universal enveloping algebra of
the algebra of differential (difference) operators in
finite-dimensional representation. In one-dimensional case
a classification is given by algebras
$sl_2({\bold R})$ (for differential operators in ${\bold R}$)
and $sl_2({\bold R})_q$ (for finite-difference operators in  ${\bold R}$),
 $osp(2,2)$  (operators in one real and one Grassmann variable, or
equivalently, $2 \times 2$ matrix operators in ${\bold R}$) and
$gl_2 ({\bold R})_K$ ( for the operators containing the differential
operators and the parity operator). A classification of linear operators
possessing infinitely many finite-dimensional invariant subspaces with
a basis in polynomials is presented.
\end{abstract}

\tableofcontents
\newpage
This Chapter is devoted to a description of a new connection between
Lie algebras and linear differential equations.

The main idea is surprisingly easy. Let us consider a certain set of
differential operators of the first order
\begin{equation}
\label{e0}
J^{\alpha}(x)\ =\ a^{\alpha , \mu}(x) \partial _{\mu} \ +
\ b^{\alpha}(x) \quad,\quad  \partial _{\mu} \equiv {d \over dx^{\mu}}
\end{equation}
$\alpha =1,2,\dots ,k \ , x \in R^{n} \ , \mu = 1,2,\dots, n$
and \ $a^{\alpha , \mu}(x) , \ b^{\alpha}(x)$ are certain functions on
$R^{n}$. Then assume that the operators form a basis of some Lie algebra $g$
of the dimension $k=\dim g$. Now take a polynomial $h$ in generators
$J^{\alpha}(x)$ and ask a question:
\begin{quote}
Does the differential operator $h(J^{\alpha}(x))$ have some specific
properties,
which distinguish this operator from a general linear differential operator?
\end{quote}
Generically, an answer is {\it negative.} However, if the algebra $g$ is taken
in a  finite-dimensional representation, the answer becomes {\it  positive} :
\begin{quote}
The differential operator $h(J^{\alpha}(x))$ does possess a finite-dimensional
invariant subspace coinciding a representation space of the finite-dimensional
representation of the algebra $g$ of differential operators of the first order.
If a basis of this finite-dimensional representation space can be constructed
explicitly, the operator $h$ can be presented in the explicit block-diagonal
form.
\end{quote}
Such differential operators having a finite-dimensional invariant subspace with
an explicit basis in functions can be named {\it quasi-exactly-solvable}.

Up to our knowledge and understanding the first {\it explicit} examples
of quasi-exactly-solvable problems were found by Razavy \cite{raz1,raz2}
and by Singh-Rampal-Biswas-Datta \cite{srbd}.
In an explicit form a general idea of quasi-exactly-solvability had been
formulated
at the first time in Turbiner \cite{t1,t11} and it had led to a complete
catalogue of
one-dimensional, quasi-exactly-solvable Schroedinger operators in connection
with
spaces of polynomials Turbiner \cite{t2}. The term \lqq
quasi-exactly-solvability" has been
suggested in Turbiner-Ushveridze \cite{tu}. A connection of
quasi-exactly-solvability
and finite-dimensional representations of  $sl_2$ was mentioned at the first
time
by Zaslavskii-Ulyanov \cite{zu}.
Later, the idea of quasi-exactly-solvability was generalized to
multidimensional
differential operators,  matrix differential operators (Shifman and Turbiner
\cite{st}),
finite-difference operators (Ogievetsky and Turbiner \cite{ot}) and, recently,
to \lqq mixed" operators containing differential operators and permutation
operators
(Turbiner \cite{tca}).

In Morozov et al \cite{mprst} it was described a connection of
quasi-exactly-solvability
with conformal quantum field theories (see also Halpern-Kiritsis \cite{hk}),
while
recent development can be found in Tseytlin \cite{tse} and also in a brief
review done
by Shifman \cite{shi}. Relationship with solid-state physics is given in
(Ulyanov and Zaslavskii \cite{uz}), while recently very exciting results in
this direction
were found by Wiegmann-Zabrodin \cite{wz1,wz2}. A general survey of a
phenomenon
of the quasi-exactly-solvability was done in Turbiner \cite{tams}.

Given Chapter will be devoted mainly to a description of quasi-exactly-solvable
operators acting on functions in one real (complex ) variable.

\renewcommand{\theequation}{1.{\arabic{equation}}}
\setcounter{equation}{0}
\section{Generalities}
Let us take $n$ linearly-independent functions $f_1(x), f_2(x) \ldots f_n(x)$
and form a linear space
\begin{equation}
\label{e1.1}
{\cal F}_n (x)\ = \ \langle f_1(x), f_2(x), \dots , f_n(x) \rangle ,
\end{equation}
where $n$ is a non-negative integer and $x \in {\bold R}$.
\begin{definition} {Two spaces ${\cal F}_n^{(1)} (x)$ and
${\cal F}_n^{(2}) (x)$ are named {\it equivalent} spaces, if one space can be
obtained through another one via a change of the variable and/or the
multiplication on a some function (making a gauge transformation), }
\begin{equation}
\label{e1.2}
{\cal F}_n^{(2)} (x)= g(x){\cal F}_n^{(1)} (y(x))
\end{equation}
\end{definition}
Choosing a certain space (1.1) and considering all possible changes of the
variable,
$y(x)$ and gauge functions, $g(x)$,  one can describe a whole class of
equivalent
spaces. It allows to introduce a certain standartization: $f_1(x)=1, f_2(x)=x$,
since
in each class of spaces, one can find a representative satisfying the
standartization.
Thus, hereafter the only spaces of the following form
\begin{equation}
\label{e1.3}
{\cal F}_n (x)\ = \ \langle 1, x, f_1(x), f_2(x), \dots , f_n(x) \rangle ,
\end{equation}
are taking into account.

It is easy to see that once an operator $h(y,d_y)$ acts on a space
${\cal F}_n^{(1)} (y)$, one can construct an operator acting on an
equivalent space ${\cal F}_n^{(2)} (y(x))$
\begin{equation}
\label{e1.4}
\bar h(y,d_y)= g(x) h(x,d_x) g^{-1}(x) |_{x=x(y)} \quad , \quad d_x \equiv {d
\over dx} \ .
\end{equation}

Later it will be considered one of the most important
particular case of the space (1.3): the space of polynomials of finite order
\begin{equation}
\label{e1.5}
{\cal P}_{n+1}(x) \ = \ \langle 1, x, x^2, \dots , x^n \rangle ,
\end{equation}
where $n$ is a non-negative integer and $x \in {\bold R}$. It is worth noting
the
important property of an invariance of the space (1.5):
\begin{equation}
\label{e1.6}
x^n {\cal P}_{n+1}({1 \over x}) \ \equiv \  {\cal P}_{n+1}(x),
\end{equation}
It stems from an evident feature of polynomials: let  $p_n(x) \in  {\cal
P}_{n+1}(x) $
is any polynomial, then $x^n p_n(1/x)$ remains a polynomial with inverse
order of the coefficients then in $p_n(x)$.

\renewcommand{\theequation}{2.{\arabic{equation}}}
\setcounter{equation}{0}
\section{Ordinary differential equations}
\subsection{General consideration}
\begin{definition} Let us name a linear differential operator of the $k$th
order, $T_k(x,d_x)$ {\it quasi-exactly-solvable}, if it preserves the space
${\cal P}_{n+1}$. Correspondingly, the operator $E_k(x,d_x)$
is named {\it exactly-solvable}, if it preserves the infinite flag
$ {\cal P}_1 \subset  {\cal P}_2 \subset {\cal P}_3
\subset \dots \subset {\cal P}_n \subset \dots $
of spaces of all polynomials: $E_k(x,d_x): {\cal P}_j \mapsto {\cal P}_j \ ,\
j=0,1,\ldots $.
\end{definition}

\begin{LEMMA} (Turbiner \cite{tams})
 (i) Suppose $n > (k-1)$.  Any quasi-exactly-solvable
operator $T_k$ can be represented by a $k$-th degree polynomial of the
operators
\label{e2.1}
\[ J^+_n = x^2 d_x - n x,\  \]
\begin{equation}
 J^0_n = x d_x - {n \over 2} \  ,
\end{equation}
\[ J^-_n = d_x \ ,  \]
\noindent
 (the operators (2.1) obey the $sl_2({\bold R})$ commutation relations:
$[J^{\pm},J^0]=\pm J^{\pm}$,\linebreak  $[J^+,J^-]=-2J^0$
\footnote{The representation (2.1) is one of the \lq projectivized'
representations (see Turbiner \cite{t1,t11}). This realization of $sl_2({\bold
R})$
has been derived at the first time by Sophus Lie.}).
If $n \leq (k-1)$, the part of the quasi-exactly-solvable operator $T_k$
containing derivatives up to order $n$ can be represented by an $n$th
degree polynomial in the generators (2.1).

 (ii) Conversely, any polynomial in (2.1) is quasi-exactly solvable.

 (iii) Among quasi-exactly-solvable operators
there exist exactly-solvable operators $E_k \subset T_k$.
\end{LEMMA}

\begin{definition}  Let us name a  {\it universal enveloping algebra} $U_g$ of
a
Lie algebra $g$ the algebra of all ordered polynomials in generators
$J^{\pm,0}$.
The notion {\it ordering} means that in any monomial in generators $J^{\pm,0}$
all $J^{+}$ are placed to the left and all $J^{-}$ to the right.
\end{definition}

{\it Comment 2.1} . \  Notion the universal enveloping algebra allows to
make a statement that $T_k$ at $k < n+1$ is simply an element
of the universal enveloping algebra $U_{sl_2({\bold R})}$ of the algebra
$sl_2({\bold R})$ taken in representation (2). If $k \geq n+1$, then $T_k$ is
represented as an element of $U_{sl_2({\bold R})}$ plus $B {d^{n+1}
\over dx^{n+1}}$, where $B$ is any linear differential operator of order
not higher than $(k-n-1)$. In otherwords, the algebra of differential operators
acting on the space (1.4) coincides to the universal enveloping algebra
$U_{sl_2({\bold R})}$ of the algebra $sl_2({\bold R})$ taken in representation
(2.1) plus operators annihilating (1.4).

{\it Comment 2.2} . \  The algebra (2.1) has the following invariance property
\[
x^{-n}\ J^{\pm,0}_n(x, d_x)\ x^n \mid _ {x={1 \over z}} \ \Rightarrow
\ J^{\pm,0}_n(z, d_z)
\]
as a consequence of the invariance (1.6) of the space ${\cal P}_{n+1}$.
In particular, if $z=1/x$, then
\begin{eqnarray}
\label{e2.2}
&J^+_n(x, d_x) \Rightarrow -J^{-}_n(z, d_z)
\nonumber \\
&J^0_n(x, d_x) \Rightarrow -J^{0}_n(z, d_z)
\nonumber \\
&J^-_n(x, d_x) \Rightarrow -J^{+}_n(z, d_z)
\end{eqnarray}

Let us introduce the {\it grading} of generators (2.1) as follows. It is easy
to check that any $sl_2({\bold R})$-generator maps a monomial into monomial,
$J^{\alpha}_n x^p \mapsto x^{p+d_{\alpha}}$.
\begin{definition}
The number $d_{\alpha}$ is named a {\it grading} of  the generator
$J^{\alpha}_n$:
$deg ( J^{\alpha}_n ) = d_{\alpha} $.
\end{definition}
Following this definition
\begin{equation}
\label{e2.3}
deg (J^+_n) = +1 \ , \ deg (J^0_n) = 0 \ , \ deg (J^-_n) = -1 ,
\end{equation}
and
\begin{equation}
\label{e2.4}
deg [(J^+_n)^{n_+} (J^0_n)^{n_0}(J^-_n)^{n_-}] \  = \ n_+ - n_- .
\end{equation}
Notion of the grading allows us to classify the operators $T_k$ in the
Lie-algebraic sense.
\begin{LEMMA} {\it A quasi-exactly-solvable operator $T_k \subset
U_{sl_2({\bold R})}$ has no terms of positive grading, if and only if
it is an exactly-solvable operator.}
\end{LEMMA}

\noindent
{\it Comment 2.3} .\  Any exactly-solvable operator having term of negative
grading
possesses terms of positive grading after transformation (2.2).  A
quasi-exactly-solvable
operator always possesses terms of positive grading as in $x$-space
representation, as in $z$-space representation.

\begin{THEOREM}  (Turbiner \cite{tams}) Let $n$ be a non-negative integer. Take
the eigenvalue
problem for a linear differential
operator of the $k$th order in one variable
\begin{equation}
\label{e2.5}
 T_k(x,d_x) \varphi (x)\ = \ \varepsilon \varphi (x)\ ,
\end{equation}
where $T_k$ is symmetric. The problem (2.5) has $(n+1)$ linearly independent
eigenfunctions in the form of a polynomial in variable $x$ of order
not higher than $n$, if and only if $T_k$ is quasi-exactly-solvable.
The problem (2.5) has an infinite sequence of polynomial eigenfunctions,
if and only if the operator is exactly-solvable.
\end{THEOREM}

\noindent
{\it Comment 2.4} .\ The \lqqp if "
 part of the first statement is obvious.
The \lqqp only if " part is a direct corollary of Lemma 2.1 .

This theorem gives a general classification of differential equations
\begin{equation}
\label{e2.6}
 \sum_{j=0}^{k} a_j (x) d_x^j \varphi (x) \ = \ \varepsilon \varphi(x)
\end{equation}
having at least one polynomial solution in $x$.
The coefficient functions $a_j (x)$ must have the form
\begin{equation}
\label{e2.7}
a_j (x) \ = \ \sum_{i=0}^{k+j} a_{j,i} x^i
\end{equation}
The explicit expressions (2.7) for coefficient function in (2.6) are obtained
by the substitution (2.1) into a general, $k$th degree polynomial element of
the
universal enveloping algebra $U_{sl_2({\bold R})}$.  Thus the coefficients
$a_{j,i}$ can be expressed  through the coefficients of the
$k$th degree polynomial element of the universal
enveloping algebra $U_{sl_2({\bold R})}$. The number of free parameters of the
polynomial solutions is defined by the number of parameters
characterizing a general $k$th degree polynomial element of the universal
enveloping algebra $U_{sl_2({\bold R})}$. A rather straightforward calculation
leads to the following formula
\begin{equation}
\label{e2.8}
 par (T_k) = (k+1)^2
\end{equation}
where we denote the number of free parameters of operator $T_k$ by the symbol
$par(T_k)$.
For the case of an infinite sequence of polynomial solutions expression
(2.7) simplifies to
\begin{equation}
\label{e2.9}
a_j (x) \ = \ \sum_{i=0}^{j} a_{j,i} x^i
\end{equation}
in agreement with the results by Krall  \cite{kr} (see also Littlejohn
\cite{little}).
In this case the number of free parameters is equal to
\begin{equation}
\label{e2.10}
 par (E_k) = {(k+1)(k+2) \over 2}
\end{equation}
On can show that the operators $T_k$ with the coefficients (2.9) preserve
a finite flag  $ {\cal P}_0 \subset {\cal P}_1 \subset {\cal P}_2 \subset \dots
\subset {\cal P}_k $ of spaces of polynomials. One can easily verify that the
preservation of such a finite flag of spaces of polynomial  leads to the
preservation
of an infinite flag of such spaces.

A class of spaces equivalent to the space of polynomials (1.5) is presented by
\begin{equation}
\label{e2.11}
\langle \alpha (z), \alpha (z) \beta (z), \dots , \alpha (z) \beta (z) ^n
\rangle \ ,
\end{equation}
where $\alpha (z),\beta (z)$ are any functions. Linear differential
operators acting in the (2.11)  are easily obtained from the
quasi-exactly-solvable
operators (2.5)--(2.7) (see Lemma 2.1)
making the change of variable $x = \beta (z)$ and the \lqq gauge"
transformation $\tilde T = \alpha (z) T  \alpha (z)^{-1}$ and have the form
\begin{equation}
\label{e2.12}
 \bar{T}_k\  = \ \alpha (z) \sum_{j=0}^{k}  (\sum_{i=0}^{k+j} a_{j,i} \beta
(z)^i )
(d^j_x)\mid_{x = \beta (z)}\alpha (z) ^{-1}
\end{equation}
where the coefficients $a_{j,i}$ are the same as in (2.7).

So the expression (2.12) gives a general form of the linear differential
operator
of the $k$th order acting a space equivalent to (1.5).
Since any one- or two-dimensional invariant sub-space can be presented in
the form (2.11) and can be reduced to (1.5), the general statement
takes place:
\begin{THEOREM}
There are no linear operators possessing one- or two-dimensional
invariant sub-space with an explicit basis other than given by Lemma 2.1.
\end{THEOREM}

Therefore an eigenvalue problem (2.5), for which one eigenfunction
can be found in an explicit form, is related to the operator
 \begin{equation}
\label{e2.13}
{\bold T^{(1)}}\ =\ B(x,d_x) d_x \ +\ q_0
\end{equation}
 and its
all  modifications (1.4), occurred after a change of the variable
and a \lqq gauge" transformation.
A general differential operator possessing
two eigenfunctions in an explicit form is presented as
\begin{equation}
\label{e2.14}
{\bold T^{(2)}}\ =\ B(x,d_x) d_x ^2\ +\ q_{2} (x) d_x  \  +\ q_{1}(x),
\end{equation}
plus its all modifications owing to a change of the variable
and a \lqq gauge" transformation. Here $ B(x,d_x) $ is any linear differential
operator,
$q_0 \in {\bold R}$ and  $q_{1,2} (x)$ are the first- and second-order
polynomials,
respectively, with coefficients such that  $q_{2} (x) d_x +q_{1}(x) $ can be
expressed
as a linear combination of the generators  $ J_1^{\pm,0}$ (see (2.1)).
\subsection{Second-order differential equations}

The second-order differential equations play exceptionally important role
in applications. Therefore, let us consider in details the second-order
differential equation (2.5) possessing polynomial solutions. From Theorem 2.1
it follows that the corresponding differential
operator must be quasi-exactly-solvable and can be represented as
\[ T_2 =  c_{++} J^+_n J^+_n  + c_{+0} J^+_n  J^0_n  + c_{+-} J^+_n  J^-_n  +
c_{0-} J^0_n  J^-_n  + c_{--} J^-_n  J^-_n  + \]
\renewcommand{\theequation}{2.15.{\arabic{equation}}}
\setcounter{equation}{0}
\begin{equation}
\label{e2.15.1}
 c_+ J^+_n  + c_0 J^0_n  + c_- J^-_n  + c ,
\end{equation}
where $c_{\alpha \beta}, c_{\alpha}, c \in {\bf R}$.
The number of free parameters is $par (T_2) = 9$. Under the condition
$c_{++}  = c_{+0}  = c_+  =0$, the operator $T_2$ becomes
exactly-solvable (see Lemma 2.2)
\begin{equation}
\label{e2.15.2}
E_2 =   c_{+-} J^+_n  J^-_n  + c_{0-} J^0_n  J^-_n  + c_{--} J^-_n  J^-_n  +
 c_0 J^0_n  + c_- J^-_n  + c ,
\end{equation}
and the number of free parameters is
reduced to $par (E_2) = 6$.

\renewcommand{\theequation}{2.{\arabic{equation}}}
\setcounter{equation}{15}

\begin{LEMMA} {\it If the operator (2.15.1) is such that
\begin{equation}
\label{e2.16}
c_{++}=0 \quad and \quad c_{+} = ({n \over 2} - m)  c_{+0} \ , \ at \ some
\ m=0,1,2,\dots
\end{equation}
 then the operator $T_2$ preserves both ${\cal P}_{n+1}$ and
${\cal P}_{m+1}$. In this case the number of free parameters is
$par (T_2) = 7$.}
\end{LEMMA}

In fact, Lemma 2.3 claims that $T_2 (J^{\alpha}_n,c_{\alpha \beta},
 c_{\alpha})$ can be rewritten as
$T_2 (J^{\alpha}_m,\allowbreak c'_{\alpha \beta},c'_{\alpha})$.
As a consequence of Lemma 2.3 and Theorem 2.1, in general,
among polynomial solutions of (2.6) there
are polynomials of order $n$ and order $m$.

{\bf Remark.}  From the Lie-algebraic point of view Lemma 2.3 means
the existence of representations of second-degree polynomials in the
generators (2.1) possessing two invariant sub-spaces.
In general, if $n$ in (2.1) is a non-negative integer, then
 among representations of $k$th degree polynomials in the generators (2.1),
lying in the universal enveloping algebra, there exist representations
possessing $1,2,...,k$ invariant sub-spaces. Even starting from an
infinite-dimensional representation of the original algebra
($n$ in (2.1) is {\it not} a non-negative integer), one can construct
the elements of the universal enveloping algebra having finite-dimensional
representation (e.g., the parameter $n$ in (2.16) is non-integer,
however $T_2$ has the invariant sub-space of dimension $(m+1)$).
Also this property implies the existence of representations of
the polynomial elements of the universal enveloping algebra
$U_{sl_2({\bold R})}$, which can be obtained starting from different
representations of the original algebra (2.1).

Substituting (2.1) into (2.15.1) and then into (2.6), we obtain
\begin{equation}
\label{e2.17}
- P_{4}(x) d_x ^2 \varphi (x) \ +\ P_{3}(x) d_x  \varphi (x) \
+\ P_{2}(x) \varphi (x) \ =\ \varepsilon \varphi (x) ,
\end{equation}
where the $P_{j}(x)$ are polynomials of $j$th order with coefficients
related to  $ c_{\alpha \beta}, c_{\alpha}$ and $n$.
In general, problem (2.17) has $(n+1)$ polynomial solutions of the form of
polynomials in $x$ of $n$th degree. If $n=1$, as a
consequence of Theorem 2.2, a  more
general spectral problem than (2.17) arises (cf. (2.14))
\begin{equation}
\label{e2.18}
- F_{3}(x) d_x ^2 \varphi (x) \ +\ q_{2}(x) d_x  \varphi (x) \
+\ q_{1}(x) \varphi (x) \ =\ \varepsilon \varphi (x) ,
\end{equation}
possessing only two polynomial solutions of the
form $(ax+b)$.
Here $F_3$ is an arbitrary complex function of $x$ and $q_j (x), j=1,2$ are
polynomials of order $j$ the same as in (2.14). For the case $n=0$ (one
polynomial
solution, $\varphi = const $) the spectral problem (2.5) becomes (cf. (2.13))
\begin{equation}
\label{e2.19}
- F_{2}(x) d_x ^2 \varphi (x) \ +\ F_{1}(x) d_x  \varphi (x) \
+\ q_0 \varphi (x) \ =\ \varepsilon \varphi (x) ,
\end{equation}
where $F_{2,1}(x)$ are arbitrary complex functions and $q_0 \in R$.

Substituting (2.1) into (2.15.2) and then into (2.6), we obtain
\begin{equation}
\label{e2.20}
- Q_{2}(x) d_x ^2 \varphi (x) \ +\ Q_{1}(x) d_x  \varphi (x) \
+\ Q_{0}(x) \varphi (x) \ =\ \varepsilon \varphi (x) ,
\end{equation}
where the $Q_{j}(x)$ are polynomials of $j$th order with arbitrary
coefficients.
One can easily show that the differential operator in the r.h.s. can be always
presented in the form (2.15.2). The  coefficients of  $Q_{j}(x)$ are
unambiguously
related to  $ c_{\alpha \beta}, c_{\alpha}$ for {\it any} value of the
parameter $n$.
Thus $n$ is a fictitious parameter and, for instance, it can be put equal to
zero.
\subsection{Quasi-exactly-solvable Schroedinger equations (examples).}
{}From the point of applications the Schroedinger equation
\begin{equation}
\label{e2.21}
(- d_z ^2 \ + \ V(z) ) \Psi (z) \ = \ \varepsilon \Psi (z),
\end{equation}
is one of the important  among the second-order differential equations (see
e.g. Landau
and Lifschitz \cite{ll}).  Here $\epsilon$ is the spectral parameter and $ \Psi
(z)$ must be
square-integrable function on some space. It can be the whole real line,
semi-line or a
finite interval. Therefore it is quite natural to search the
quasi-exactly-solvable
and exactly-solvable operators of the Schroedinger type acting on a
finite-dimensional
linear space of square-integrable functions:$f_i(z), i=1,2,\ldots n$.

One  possible way to get  the quasi-exactly-solvable and exactly-solvable
Schroedinger operators is to transform the quasi-exactly-solvable $T_2$ and
exactly-solvable $E_2$ operators acting on finite-dimensional spaces of
polynomials
(1.5) into the Schroedinger type operators. It always can be done making a
change
of a variable and a gauge transformation (see (1.2)) as a consequence of
one-dimensional nature of equations studied.
In practice, the realization of this transformation is nothing but a conversion
of (2.17)-(2.20) into (2.21). The only open question remains: does new basis
belong to square-integrable one or not? This question will be discussed in
the end of this Section. In the following consideration we restrict ourself by
the case of real functions and real variables.

Introducing a new function
\begin{equation}
\label{e2.22}
\Psi (z) \ =\ \varphi (x(z)) e ^ {-A(z)} ,
\end{equation}
and new variable $x=x(z)$ , where $A(z)$
is a certain real function, one can reduce (2.17)--(2.20) to the
Sturm-Liouville-type problem (2.21) with the potential equal to
\begin{equation}
\label{e2.23}
 V(z) = (A')^2 - A'' + P_2 (x(z)) \ ,
\end{equation}
if
\[ A = \int ({P_3 \over P_4})dx - log z'\ ,
\ z = \pm\int {dx \over \sqrt{P_4}}\ .\]
for the case of (2.17), or
\[ A = \int ({Q_2 \over F_3})dx - log z'\ ,
\ z = \pm\int {dx \over \sqrt{F_3}}\ .\]
for the case of (2.18), or
\[ A = \int ({F_1 \over F_2})dx - log z'\ ,
\ z = \pm\int {dx \over \sqrt{F_2}}\ .\]
for the case of (2.19) with replacement of  $P_2 (x(z))$ for two latter cases
by
$Q_1 (x(z))$  or $Q_0$ , respectively.
 If the functions (2.22), obtained after transformation,
belong to the ${\cal L}_2(\cal D)$-space\footnote{Depending on the
change of variable  $x = x(z)$, the space $\cal D$ can be whole
real line, semi-line and a finite interval.} ,
we arrive at the recently discovered quasi-exactly-solvable Schroedinger
equations
(Turbiner \cite{t1,t11,t2}), where a finite number of eigenstates is found
algebraically.

In order to proceed a description of  concrete examples of
quasi-exactly-solvable
Schroedinger equations, first of all let us generalize the eigenvalue problem
(2.21)
inserting a weight function $\hat \varrho(z) \equiv \varrho (x(z))$ in the
r.h.s.
\begin{equation}
\label{e2.24}
(- d_z ^2 \ + \ V(z) ) \Psi (z) \ = \ \varepsilon \varrho (x(z)) \Psi (z),
\end{equation}
where for a sake of future convenience the weight function is presented in the
form of composition $\varrho (x(z))$.
This equation can obtained from (2.17)-(2.19) by taking the same gauge factor
(2.22) as before but with another change of the variable
\[
z\ =\  \pm \int {dx \over \sqrt{\varrho (x) P_4 (x)}}\
\]
what leads to a slightly modified potential then in (2.23)
\[
V(z) = (A')^2 - A'' + P_2 (x(z)) \varrho (x(z))
\]

Below we will follow the catalogue given at  Turbiner  \cite{t2}. A
presentation of results is
the following: firstly, we display the quadratic element $T_2$ of the universal
enveloping
algebra $sl_2$ in the representation (2.1) and its
equivalent form of differential operator $T_2(x,d_2)$,
secondly, the corresponding potential $V(z)$and afterwards,
the explicit expression for  the change of the variable $x=x(z)$,
the weight function $\varrho (z) $and finally the
functional form of the eigenfunctions $\Psi (z)$ of the \lqq algebraized"
part of the spectra \footnote{The functions $p_n(x)$ occurring in the
forthcoming
expressions for $\Psi (z)$ are polynomials of the $n$th order. They are nothing
but the polynomial eigenfunctions of the operator $T_2(x,d_x)$ .}.

We begin a consideration with the quasi-exactly-solvable equations, associated
to the
exactly-solvable Morse oscillator; it implies that at the limit, when number of
\lqq algebraized" eigenstates $(n+1)$ goes to infinity, the Morse oscillator
occurs.
\vskip .5truecm
{\it Comment 2.5} \ The Morse oscillator is one of well-known exactly-solvable
quantum-mechanical problems  (see e.g. Landau and Lifschitz \cite{ll}).
It is described by the Schroedinger operator with the potential
\[
V(z) = A^2 e^{-2\alpha z} - 2 A e^{-\alpha z}  \ ,\  A,\alpha >0 \ .
\]
This potential is used to model the interaction of the atoms in diatomic
molecules.
\vskip .5truecm

{\it I}.
\vskip .3truecm
\begin{equation}
\label{e2.25}
 T_2 = -\alpha ^2J^+_n J^-_n + 2\alpha a J^+_n - \alpha [\alpha (n+1)+ 2b]
J^0_n -
2\alpha c J^-_n
\end{equation}
\[
- {\alpha n \over 2} [\alpha (n+1) +2b]
\]
or as the differential operator,
\[
T_2(x,d_x)= -\alpha ^2 x^2d_x^2 + \alpha [2ax^2-(\alpha+2b)x-2c]d_x -
2\alpha anx
\]
leads to
\begin{equation}
\label{e2.26}
 V(z) = a^2e^{-2\alpha z} - a[2b + \alpha (2n +1)]e^{-\alpha z}+
c(2b-\alpha)e^{\alpha z} + c^2e^{2\alpha z}
\end{equation}
\[
+b^2 - 2ac
\]
where
\[
 x=e^{-\alpha z}\ , \ \varrho=1 \ ,
\]
with  the eigenfunctions of  \lqq algebraized" part of the spectra
\[
\Psi (z) \ = \  p_n (e^{-\alpha z})
\exp{(-{a \over  \alpha} e^{-\alpha z} -bz - {c \over \alpha } e^{ \alpha z})}
\]
at $\alpha >0,\ a, c \geq 0$ and $\forall b$.
\vskip .5truecm

{\it II}.
\vskip .3truecm
\begin{equation}
\label{e2.27}
 T_2 = -\alpha J^0_n J^-_n + 2a J^+_n - 2c J^0_n - [\alpha ({n\over 2}+1)+ 2b]
J^-_n - cn
\end{equation}
or as the differential operator,
\[
T_2(x,d_x)= -\alpha x d_x^2 + (-2ax^2+2cx+\alpha +2b)d_x +2anx
\]
leads to
\begin{equation}
\label{e2.28}
 V(z) = a^2e^{-4\alpha z} - 2ac e^{-3\alpha z}
+ [c^2 - 2a(b +\alpha a n + \alpha)]e^{-2\alpha z} + c(2b+\alpha )e^{-\alpha z}
+b^2\ ,
\end{equation}
where
\[
 x=e^{-\alpha z}\ , \ \varrho={1 \over \alpha} e^{-\alpha z} \ ,
\]
\[
\Psi (z) \ = \  p_n (e^{-\alpha z}) \exp{(-{a \over 2\alpha }e^{-2\alpha z} +
{c \over \alpha }e^{-\alpha z} - bz )}
\]
at  $\alpha >0, a \geq 0, b >0$ and $\forall c$.
\vskip .5truecm

{\it III}.
\vskip .3truecm
\begin{equation}
\label{e2.29}
 T_2 = -\alpha J^+_n J^0_n + (2b-3\alpha n /2 ) J^+_n - 2a J^0_n - 2c J^-_n -
an
\end{equation}
or as the differential operator,
\[
T_2(x,d_x)= -\alpha x^3 d_x^2 + [(2b-\alpha)x^2-2ax-2c]d_x + (\alpha n -2b)nx
\]
leads to
\begin{equation}
\label{e2.30}
 V(z) = c^2e^{4\alpha z} +2ac e^{3\alpha z}+ [a^2 - 2c(b +\alpha)]e^{2\alpha
z}-
a(2b+\alpha) e^{\alpha z}\ +b^2+ \alpha n(\alpha n -2b) \ ,
\end{equation}
where
\[
x=e^{-\alpha z}\ ,\ \varrho={1 \over \alpha} e^{\alpha z} \ ,
\]
\[
\Psi (z) \ = \  p_n (e^{-\alpha z}) \exp{(-{c \over 2\alpha} e^{2\alpha z} -
{a \over \alpha} e^{\alpha z} + bz )}
\]
at $\alpha > 0, b >0, c\geq 0$ and $\forall a$.
\vskip .5truecm

Next two quasi-exactly-solvable potentials are associated to the one-soliton or
P\"{o}schle-Teller potential.
\vskip .5truecm

{\it Comment 2.6} \ The P\"{o}schle-Teller or one-soliton potential describes
another
well-known exactly-solvable quantum-mechanical problem  (see e.g. Landau and
Lifschitz \cite{ll}). It is given by the Schroedinger operator with the
potential
\[
V(z) = - A^2  {1 \over \cosh^2 {\alpha z}} \  .
\]
This potential has a unique property of an absence of reflection and is also
one of the
simplest solutions of so called Korteweg-de Vries equation playing important
role
in the {\it inverse problem method} (for detailed discussion see e.g. the book
by
V.E. Zakharov et al \cite{zmnp}).
\vskip .5truecm

{\it IV}.
\vskip .3truecm
\[
 T_2 = - 4 \alpha ^2 J^+_n J^0_n  + 4 \alpha ^2 J^+_n J^-_n
-2\alpha [\alpha (3n+2k+1)+2c] J^+_n
\]
\begin{equation}
\label{e2.31}
+ 2 \alpha [\alpha (n+2)+2c-2a] J^0_n + 4\alpha a J^-_n +
\alpha n[\alpha (n+2)+2c-2a]
\end{equation}
or as the differential operator,
\[
T_2(x,d_x)= -4\alpha ^2 (x^3-x^2) d_x^2 -
2\alpha [(3 \alpha + 2\alpha k+ 2c) x^2 -2 ( \alpha - a + c)x -2a]d_x
\]
\[
 + (2n+k)\alpha   [(2n+k+1)\alpha +2c] x
\]
leads to
\begin{equation}
\label{e2.32}
 V(z) = a^2 \cosh^4 {\alpha z} - a(a+2\alpha - 2c) \cosh^2 {\alpha z}
\end{equation}
\[
- [c(c+\alpha) + \alpha (2n+k)(\alpha (2n+k) + \alpha +2c)] \cosh ^{-2}
{\alpha z}
+ c^2 + a \alpha - 2ac
\]
where
\[
x=\cosh{\alpha z}^{-2}\ ,\ \varrho=1  \ ,
\]
\[
\Psi (z) \ = \  (\tanh {\alpha z})^k p_n (\tanh^2 {\alpha z}) ( \cosh{\alpha
z})^{-c/\alpha}
\exp{(-{a \over {4\alpha}} \cosh{ 2 \alpha z}  )}
\]
at  $\alpha >0, a \geq 0, \forall c$ and $k=0,1$.
\vskip .5truecm

{\it V}.
\vskip .3truecm
\[
T_2 = -4 \alpha^2 J^+_n J^-_n + 4 \alpha ^2 J^0_n J^-_n + 4\alpha b J^+_n
-2\alpha [\alpha (2n+ 2k+3)+2a+ 4b] J^0_n
\]
\begin{equation}
\label{e2.33}
+ 2\alpha [\alpha (n+2)+2a+2b) J^-_n - \alpha  [\alpha n (2n+ 2k+3)+2an - 2 b
k]
\end{equation}
or as the differential operator,
\[
T_2(x,d_x)= -4\alpha^2 x(x-1) d_x^2 +
2\alpha [2bx^2-(2a+4b+ 2k \alpha+ 3\alpha) x+ 2(\alpha+a+b)]d_x
\]
\[
- 4 \alpha bn x + 2\alpha b (2n+k)
\]
leads to
\begin{equation}
\label{e2.34}
 V(z) = -b^2 \cosh^{-6} {\alpha z} + b[2a+ 3b +\alpha (4n+2k +3)]
\cosh^{-4}{\alpha z}
\end{equation}
\[
-[(a+3b)(a+b + \alpha) +2 (2 n+k)\alpha  b )] \cosh^{-2} {\alpha z}
+ (a+b)^2
\]
where
\[
x=\cosh{\alpha z}^{-2}\ ,\ \varrho= \cosh^{-2}{\alpha z}\  ,
\]
\[
\Psi (z) \ = \  (\tanh {\alpha z})^k  p_n (\tanh^2 {\alpha z}) ( \cosh{\alpha
z})^
{-{(a+b)} \over \alpha} \exp{({b \over {2\alpha} } \tanh^2 {2\alpha z} )}
\]
at $\alpha > 0, (a +b) > 0 $ and $k=0,1$.
\vskip .5truecm
Next two quasi-exactly-solvable potentials are associated to the harmonic
oscillator potential.
\vskip .5truecm

{\it VI}.
\vskip .3truecm
It is the first example of  the quasi-exactly-solvable Schroedinger operator.
Let us take the following non-linear combination in the generators (2.1)
(Turbiner \cite{t2})
\begin{equation}
\label{e2.35}
 T_2 = -4 J^0_n J^-_n + 4a J^+_n + 4b J^0_n - 2(n+1+2k) J^-_n + 2bn
\end{equation}
or as  the differential operator,
\[
T_2(x,d_x)= -4xd_x^2 + 2(2ax^2+2bx-1-2k)d_x - 4anx  \ ,
\]
where $x \in R$ and $a>0, \forall b$ or $a \geq 0, b>0$ . Putting $x=z^2$ and
choosing the gauge phase $A=ax^2/4 + bx/2 - k/2 \ln{x}$, we arrive at the the
spectral problem (2.19)  with the potential (Turbiner and Ushveridze \cite{tu})
\begin{equation}
\label{e2.36}
 V(z) = a^2z^6 + 2abz^4 + [b^2 - (4n+3+2k)a]z^2 - b(1+2k),
\end{equation}
for which at $k=0\ (k=1)$ the first $(n+1)$ eigenfunctions, even (odd)
in $x$, can be found algebraically. Of course, the number of those
\lqq algebraized" eigenfunctions is nothing but the dimension of the
irreducible representation (1.4) of the algebra (2.1).  $(n+1)$ `algebraized'
eigenfunctions of (2.19) have the form
\begin{equation}
\label{e2.37}
\Psi (z) \ =\ z^k p_n (z^2)  e ^ {-{az^4 \over 4} - {bz^2 \over 2}},
\end{equation}
where $p_n(y)$ is a polynomial of the $n$th degree. It is worth noting that
if the parameter $a$ goes to 0, the potential (2.35) becomes the harmonic
oscillator potential and polynomials $z^k p_n (z^2)$ reconcile to the Hermite
polynomials  $H_{2n+k} (z)$ (see discussion below).
\vskip .5truecm
{\it VII}.
\vskip .3truecm
\begin{equation}
\label{e2.38}
 T_2 =
 -4 J^0_n J^-_n + 4a J^+_n + 4b J^0_n - 2(n+d+2l-2c) J^-_n + 2bn
\end{equation}
or as the differential operator,
\[
T_2(x,d_x)= -4xd_x^2 + 2(2ax^2+2bx - d - 2l + 2c)d_x - 4anx
\]
leads to
\begin{equation}
\label{e2.39}
  V(z) = a^2z^6 + 2abz^4 + [b^2 - (4n+2l+d+2-2c)a]z^2 +c(c-2l-d+2)z^{-2}
\end{equation}
\[
-b(d+2l-2c),
\]
where
\[
x=z^2\ ,\ \varrho=1  \ ,
\]
\[
\Psi (z) \ = \  p_n (z^2) z^{l-c}  e ^ {-{az^4 \over 4} - {bz^2 \over 2}},
\]
at $a>0, \forall b$ or $a \geq 0, b>0$ and $(d+l-c)>1$ . This case corresponds
to the radial part of $d$-dimensional Schroedinger equation, $z \in [0, \infty
)$.
At $d=1$ the radial part coincides with the ordinary Schroedinger equation
(2.19)
at $z \in R$. The potential (2.39) becomes a generalized version of the
potential (2.36)
with an additional singular term proportional to $ z^{-2}$.
\vskip .5truecm

Next two quasi-exactly-solvable potentials are associated to the Coulomb
problem.
\vskip .5truecm

{\it VIII}.
\vskip .3truecm
\begin{equation}
\label{e2.40}
 T_2 =
 -J^0_n J^-_n + 2a J^+_n + 2b J^0_n - (n/2+d+2l-2c-1) J^-_n - an
\end{equation}
or as the differential operator,
\[
T_2(x,d_x)= -xd_x^2 + (2ax^2+2bx+2c-d-2l+1)d_x - 2anx
\]
leads to
\begin{equation}
\label{e2.41}
  V(z) = a^2z^2 + 2abz - b (2l+d-1-2c)z^{-1} +c(c-2l-d+2)z^{-2}
\end{equation}
\[
+b^2 + a (2c -d- 2l -2n) \ ,
\]
where
\[
x=z\ ,\ \varrho=z^{-1}  \ ,
\]
\[
\Psi (z) \ = \  p_n (z) z^{l-c}  e ^ {-{a \over 2} z^2  - bz},
\]
at $ a \geq 0, b>0$ and $(d+l-c)>1$.
\vskip .5truecm
{\it IX}.
\vskip .3truecm
\begin{equation}
\label{e2.42}
 T_2 =
 -J^+_n J^-_n + 2a J^+_n - (n+d-1+2l-2c) J^0_n + 2b J^-_n -n (d+2l-1-2c)
\end{equation}
or as the differential operator,
\[
T_2(x,d_x)= -x^2d_x^2 - [2ax^2+(2c-d-2l+1)x -2b]d_x - 4anx
\]
leads to
\begin{equation}
\label{e2.43}
  V(z) = b^2z^{-4} - b(2c-2l-d+3)z^{-3} + [c (c-2l-d+2)-2ab]z^{-2}
\end{equation}
\[
-a(2n+2l+d-1-2c)z^{-1} + a^2\ ,
\]
where
\[
x=z\ ,\ \varrho=z^{-2}  \ ,
\]
\[
\Psi (z) \ = \  p_n (z) z^{l-c}  e ^ { -az - bz^{-1} },
\]
at $a \geq 0, b>0$ or $a >0, b \geq 0$ and  $c, l, d  \in R$.
\vskip .5truecm

Now let us show an example of the non-singular periodic quasi-exactly-solvable
potential connected with Mathieu potential.
\vskip .5truecm
{\it Comment 2.7} \ The Mathieu potential
\[
V(z)\ =\ A \cos{\alpha z}
\]
is one of the most important potentials in many branches of physics,
engineering etc.
Detailed description of the properties of the corresponding Scroedinger
equation
can be found, for instance, in Bateman-Erd\'elyi \cite{be}, vol.3 and also
Kamke
\cite{kamke}, Equation 2.22.

\vskip .5truecm

{\it X}.
\vskip .3truecm
Firstly, take the following operator
\begin{equation}
\label{e2.44}
 T_2 =
 \alpha ^2 J^+_{n-\mu} J^-_{n-\mu}   - \alpha ^2 J^-_{n-\mu}  J^-_{n-\mu}  +
\end{equation}
\[
2a \alpha J^+_{n-\mu}  +\alpha ^2 (n+\mu+1) J^0_{n-\mu}  - 2\alpha a
J^-_{n-\mu}
+ \alpha^2 {(n-\mu)(n+\mu+1) \over 2}
\]
or as the differential operator,
\[
T_2(x,d_x)= \alpha^2 (x^2-1) d_x^2 + \alpha [2ax^2+ \alpha (1+2\mu)x-2a]d_x -
2\alpha a(n-\mu)x
\]
The transformation (2.22)-(2.23) leads to the periodic potential
\begin{equation}
\label{e2.45}
  V(z) =  a ^2  \sin^2{\alpha z} -  \alpha a (2n+1) \cos{\alpha z} + \mu
\alpha^2,
\end{equation}
where
\[
x=\cos{\alpha z} \ ,\ \varrho=1  \ ,
\]
\[
\Psi (z) \ = \ (\sin {\alpha z})^{\mu}   p_{n-\mu} (\cos{\alpha z} ) \
e ^ {( {a \over \alpha}\cos{\alpha z} )},
\]
at $ a \geq 0,  \alpha \geq 0$. Here $\mu = 0,1$ . For the fixed $n$, $(2n+1)$
eigenstates
having a meaning of the edges of bands can be found algebraically.

Secondly, take the following operator
\begin{equation}
\label{e2.46}
 T_2 =
 \alpha ^2 J^+_{n-1} J^-_{n-1}   - \alpha ^2 J^-_{n-1}  J^-_{n-1}  +
\end{equation}
\[
2a \alpha J^+_{n-1}  +\alpha ^2 (n+1) J^0_{n-1}  - \alpha [2a+\alpha
(\nu_1-\nu_2)]  J^-_{n-1}
+ \alpha^2 {(n^2-1) \over 2}
\]
or as the differential operator,
\[
T_2(x,d_x)= \alpha^2 (x^2-1) d_x^2 + \alpha [2ax^2+ 2 \alpha x-2a- \alpha
(\nu_1-\nu_2) ]d_x -
2\alpha a(n-1)x
\]
The transformation (2.22)-(2.23) leads to the periodic potential
\begin{equation}
\label{e2.47}
  V(z) =  a ^2  \sin^2{\alpha z} -  \alpha a (2n) \cos{\alpha z} + \alpha a
(\nu_1-\nu_2)]
- { \alpha^2 \over 4} \ ,
\end{equation}
where
\[
x=\cos{\alpha z} \ ,\ \varrho=1  \ ,
\]
\[
\Psi (z) \ = \ (\cos{\alpha z})^{\nu_1}  (\sin{\alpha z})^{\nu_2}   p_{n-1}
(\cos{\alpha z} ) \
e ^ {( {a \over \alpha}\cos{\alpha z} )},
\]
at $ a \geq 0,  \alpha \geq 0$. Here $\nu_{1,2} = 0,1$, but $\nu_1 + \nu_2 =1$
. For the fixed $n$, $(2n+1)$ eigenstates
having a meaning of the edges of bands can be found algebraically.

\subsection{Quasi-exactly-solvable Schroedinger equations (Lame equation).}

In this Section we consider one of the most important second-order ordinary
differential equation -- $m$-zone Lame equation
\begin{equation}
\label{e2.4.1}
- d_x^2 \Psi + m(m + 1) {\cal P}(x) \Psi \ =\  \varepsilon \Psi
\end{equation}
where ${\cal P}(x)$ is the Weierstrass function in a standard
notation (see, e.g. Bateman and A. Erd\'{e}lyi \cite{be}), which depends on two
free parameters, and $m=1,2,\ldots$ .
In a description we mainly follow the paper (Turbiner \cite{tlame}).

The Weierstrass function is a double-periodic meromorphic function for which
the equation ${\cal P}^{\prime 2} = ({\cal P} - e_1)({\cal P} - e_2)({\cal P} -
e_3)$
holds, where $\sum e_i=0$.  Introducing the new variable
$\xi = {\cal P}(x)+{1\over 3} \sum a_i$ in ((\ref{e2.4.1})) (see, e.g. Kamke
\cite{kamke}), the new equation emerges
\begin{equation}
\label{e2.4.2}
	\eta^{\prime\prime} + {1\over 2} \big ({1\over \xi - a_1} +
	{1\over \xi - a_2} + {1\over \xi - a_3}\big ) \eta^\prime -
	{m(m+1) \xi + \varepsilon \over 4(\xi - a_1)(\xi - a_2)(\xi -
	a_3)} \eta = 0
\end{equation}
 where $\eta(\xi)\equiv\psi(x)$.  Here the new parameters
$a_i$ satisfy the system of linear equations $e_i = a_i - {1\over 3} \sum a_i$.
Equation (\ref{e2.4.2}) is named by an algebraic form for the Lame
equation.  There exists a spectral parameter $\lambda$ for which
equation (\ref{e2.4.2}) has four types of solutions:
\renewcommand{\theequation}{2.50.{\arabic{equation}}}
\setcounter{equation}{0}
\begin{equation}
\label{e2.4.3.1}
	\eta^{(1)} \ = \ p_k(\xi)
\end{equation}
\begin{equation}
\label{e2.4.3.2}
	\eta^{(2)}_i \ =\  (\xi - a_i)^{1/2} p_k(\xi)\quad, \quad i =1,2,3
\end{equation}
\begin{equation}
\label{e2.4.3.3}
	\eta^{(3)}_i \ = (\xi - a_{l_1}l)^{1/2} (\xi - a_{l_2})^{1/2}
	p_{k-1}(\xi)\ , \quad l_1 \ne l_2; i \ne l_{1,2}; i,l_{1,2} = 1,2,3
\end{equation}
\begin{equation}
\label{e2.4.3.4}
	\eta^{(4)} \ = (\xi - a_1)^{1/2} (\xi - a_2)^{1/2} (\xi - a_3) p_{k-1}(\xi)
\end{equation}

\renewcommand{\theequation}{2.{\arabic{equation}}}
\setcounter{equation}{50}

\noindent
where $p_r (\xi)$ are polynomial in $\xi$ of degree $r$.  If the value of
parameter $n$ is fixed, there are $(2m + 1)$ linear independent
solutions of the following form: if $m = 2k$ is even, then the
$\eta^{(1)}(\xi)$ and $\eta^{(3)}(\xi)$ solutions arise, if $m = 2k +1$ is odd
we have solutions of the $\eta^{(2)}(\xi)$ and $\eta^{(4)}(\xi)$ types.
Those eigenvalues have a meaning of the edges of the zones in the potential
(\ref{e2.4.1}).

\begin{THEOREM} (Turbiner \cite{tlame})
\  The spectral problem (\ref{e2.4.1})  at $m=1,2,\ldots$ with polynomial
solutions
(\ref{e2.4.3.1}), (\ref{e2.4.3.2}), (\ref{e2.4.3.3}), (\ref{e2.4.3.4}) is
equivalent
to the spectral problem (2.5) for the operator $T_2$ (2.15.1) belonging the
universal enveloping $sl_2$-algebra in the representation (2.1) with the
coefficients
\begin{equation}
\label{e2.4.4}
	c_{+0} = 4 \quad ,\quad c_{+-} = -4 \sum a_i \quad ,\quad c_{0-}
	= 4 \sum a_i a_j \quad ,\quad c_{--} = a_1 a_2 a_3
\end{equation}
before the terms quadratic in generators and  the following coefficients before
linear in
generators $J ^{\pm,0}_r$ :
\begin{enumerate}
\item[(1)] \ For $\eta^{(1)}(\xi)$-type solutions at $m=2k, r=k$
\renewcommand{\theequation}{2.52.{\arabic{equation}}}
\setcounter{equation}{0}
\begin{equation}
\label{e2.4.5.1}
c_+ \ = - 6k -2\ , \  c_0\ = \ 4(k + 1)\sum a_i \ ,\ c_- = -2(k + 1)\sum a_i
a_j
\end{equation}
\item[(2)] \ For $\eta^{(2)}_i (\xi)$-type solutions at $m=2k+1, r=k$
\[
c_+\ =\ -6k -6\ ,\ c_0 \ = 4(k + 2)\sum a_i  - a_i\ , \
\]
\begin{equation}
\label{e2.4.5.2}
c_-\ =\ -2(k + 1)\sum a_i a_j - 4a_{l_1} a_{l_2} \ ,\quad  i \ne l_{1,2} , l_1
\ne l_2
\end{equation}
\item[(3)] \ For $\eta^{(3)}_i (\xi)$-type solutions at $m=2k, r=k-1$
\[
c_+\ =\ -6k -4 \ ,\   c_0\ = 4(k + 1)\sum a_i  + 4a_i\ ,
\]
\begin{equation}
\label{e2.4.5.3}
\ c_-\ =\ -2(k + 2)\sum a_i a_j + 4a_{l_1} a_{l_2} \ ,\quad i \ne l_{1,2} , l_1
\ne l_2
\end{equation}
\item[(4)]  \ For $\eta^{(4)}_i (\xi)$-type solutions at $m=2k+1, r=k-1$
\begin{equation}
\label{e2.4.5.4}
c_+ = -6k - 8 \ ,\ c_0 = 4(k + 2)\sum a_i  \ ,\ c_- = -2(k + 2)\sum a_i a_j
\end{equation}
\end{enumerate}
\renewcommand{\theequation}{2.{\arabic{equation}}}
\setcounter{equation}{52}
\end{THEOREM}
So, each type of solution (\ref{e2.4.3.1}), (\ref{e2.4.3.2}),
(\ref{e2.4.3.3}), (\ref{e2.4.3.4})  corresponds to the
particular spectral problem (\ref{e2.4.1})  with a special set of parameters
(\ref{e2.4.4}) plus (\ref{e2.4.5.1}), (\ref{e2.4.5.2}), (\ref{e2.4.5.3}),
 (\ref{e2.4.5.4}), respectively.
 It can be easily shown that the calculation of eigenvalues $\varepsilon$ of
(2.48)
corresponds to the solution of a characteristic equation for the
four-diagonal matrix:
\[
         C_{l,l-1} \ =\ (l - 1 -2j) [4(j + 1 - l) + c_+] \ ,
\]
\[
	C_{l,l} \ = \ [l(2j + 1 - l)c_{+-} + (l - j) c_0] \ ,
\]
\[
	C_{l,l+1} \ = \ (l + 1)(j - l) c_{0-} + (l + 1) c_- \ ,
\]
\begin{equation}
\label{e2.4.6}
	C_{l,l+2} \ =\  - (l + 1)(l + 2) c_{--}.
\end{equation}
where the size of this matrix is $(k + 1) \times (k + 1)$ and $2j=k$ for
(\ref{e2.4.3.1}),
(\ref{e2.4.3.2}), and $k \times k$ and $2j=k-1$ for (\ref{e2.4.3.3}),
(\ref{e2.4.3.4}),
respectively.  In connection to Theorem 2.3 one prove the following theorem
 (Turbiner \cite{tlame}).

\begin{THEOREM}
\  Let fix the parameters $e'$s ($a'$s) in (\ref{e2.4.1}) (or
(\ref{e2.4.2})) except one, e.g. $e_1 (a_1)$. The first  $(2m + 1)$ eigenvalues
of
(\ref{e2.4.1}) (or  (\ref{e2.4.2})) form  $(2m + 1)$-sheeted Riemann surface
 in parameter $e_1 (a_1)$. This surface contains four disconnected pieces:
one of them corresponds to $\eta^{(1)} (\eta^{(4)})$ solutions and the
others correspond to $\eta^{(3)} (\eta^{(2)})$.
At $m = 2k$ the Riemann subsurface for $\eta^{(1)}$ has $(k + 1)$ sheets and
the
number of sheets in each of the others is equal to $k$.
At $m = 2k + 1$ the number of sheets for $\eta^{(4)}$ is equal to $k$ and for
$\eta^{(2)}$
each subsurface contains $(k + 1)$ sheets.
\end{THEOREM}

It is worth emphasizing that we cannot find a relation between
the spectral problem for the two-zone potential
\begin{equation}
\label{e2.4.7}
	V = -2 \sum^3_{k=1} {\cal P} (x - x_i) \ ,\quad \sum^3_{i=1} x_i = 0 \ ,
\end{equation}
(see Dubrovin and Novikov \cite{dn})
\footnote{The potential (\ref{e2.4.7}) and the original Lame potential
(\ref{e2.4.1})
at $m=2$ are related to via the isospectral deformation.}
and the spectral problem (2.5) for $T_2$ with the parameters (\ref{e2.4.4}) and
(\ref{e2.4.5.1}) or  (\ref{e2.4.5.3})  at $k=1$.  In this case eigenvalues
$\varepsilon$ and also eigenfunctions (2.49) (but not (2.48) do not depend
on parameters $c_{--}$.

{\it Comment 2.8}  .\
One can generalize a meaning of isospectral deformation, saying we want
to study a variety of potentials with the first several coinciding eigenvalues.
It can be named {\it quasi-isospectral deformation}.

Now let us consider such a quasi-isospectral deformation of  (\ref{e2.4.1}) at
$m=2$.
It arises from the fact that the addition of the term $c_{++} J^+_rJ^+_r$ to
the operator
$T_2$ with the parameters (\ref{e2.4.4}) and (\ref{e2.4.5.1}) or
(\ref{e2.4.5.3}) at $k=1$
does not change the characteristic matrix  (\ref{e2.4.6}).
Making the reduction (2.22)-(2.23) from the equation (2.5) to the Schroedinger
equation
(2.21), we obtain
\begin{equation}
\label{e2.4.8}
V(x) = c_{++} {2 c_{++} \xi^6 - c_{+-} \xi^4 - 2 c_{0-} \xi^3 -3 c_{--} \xi^2
\over P^2_4 (\xi)} +  P_2 (\xi) \ ,
\end{equation}
where
\begin{equation}
\label{e2.4.9}
P_4 (\xi) =c_{++} \xi^4 + c_{+0} \xi^3 + c_{+-} \xi^2 + c_{0-} \xi + c_{--} \
,\
P_2 (\xi) = - m (m + 1) \xi  + {c_0\over 2}\ .
\end{equation}
and $\xi$ is defined via the equation
\begin{equation}
\label{e2.4.10}
	x = \int {d\xi \over \sqrt{P_4 (\xi)}} \ ,
\end{equation}
In general, the potential (\ref{e2.4.8}) contains four double poles in $x$ and
does not reduce to (\ref{e2.4.7}).  It is worth noting that the first five
eigenfunctions
in the potential (\ref{e2.4.8}) have the form
\begin{equation}
\label{e2.4.11}
	\Psi (x) =
\left\{ \begin{array}{c}
A\xi + B  \\ (\xi - a_i)^{1/2} (\xi - a_j)^{1/2}
\end{array}  \right\}
\exp { \left( -c_{++} \int {\xi^3 d\xi\over P_4 (\xi)}  \right) } \ ,\ i\ne j,\
i,j =1,2,3 .
\end{equation}
Here $\xi$ is given by (\ref{e2.4.10}).  The first five eigenvalues of the
potential (\ref{e2.4.8}) do not depend on the parameters $c_{--}, c_{++}$.

\subsection{Exactly-solvable equations and classical orthogonal polynomials.}

As one of the most important properties of the exactly-solvable operators
(2.15.2)
 is the following: the eigenvalues of a general exactly-solvable operator $E_2$
are
given by the quadratic polynomial in number of eigenstate
\begin{equation}
\label{e2.5.1}
{\epsilon}_m =  \ c_{00} m^2 \ + \ c_{0} m \ + \ const
\end{equation}
(for details see  Turbiner \cite{t1,t11}). It can be easily verified by
straightforward
calculation.

Taking different exactly-solvable operators $E_2$ (see (2.15.2)) for the
eigenvalue
problem (2.5) one can reproduce the equations having the Hermite, Laguerre,
Legendre and Jacobi polynomials as eigenfunctions (Turbiner \cite{t1,t11}),
which
is shown below.  In the definition of the about polynomials we follow the
definition
given in (Bateman and A. Erd\'{e}lyi \cite{be}).

{\it1. Hermite polynomials.}

The Hermite polynomials $H_{2m+k},\ m=0,1,2,\ldots , \ k=0,1$ are the
polynomial
eigenfunctions of the operator
\begin{equation}
\label{e2.5.2}
E_2 (x, d_x) = d_x^2 - 2x d_x
\end{equation}
which immediately can be rewritten in terms of the generators (2.1) following
the
Lemma 2.1
\begin{equation}
\label{e2.5.3}
E_2 = \ J^-_0 (x)J^-_0 (x)\ -\ 2 J^0_0 (x) \
\end{equation}
However, there exists another way to represent the operators related to the
Hermite
polynomials. Let us notice that $k$ has a meaning of the parity of the
polynomial
$H_{2m+k}$ and
\[
H_{2m+k} (x) = x^k h_m (x^2)
\]
Then it is easy to find that the operator having $h_m(y)$ as the eigenfunctions
\begin{equation}
\label{e2.5.4}
\bar{E}_2 (y, d_y) = 4yd_y^2 - 2(2y-1-2k) d_y
\end{equation}
and, correspondingly
 \begin{equation}
\label{e2.5.5}
\bar{E}_2 = \ 4J^0_0 (y)J^-_0 (y)\ -\ 4 J^0_0 (y) \ +\ 2(1+2k) J^-_0 (y)
\end{equation}
Of course, those two representations are equivalent, however, a
quasi-exactly-solvable
generalization can be implemented for the second representation only
(see examples VI and VII in Section 2.3).

{\it 2. Laguerre polynomials.}

The associated Laguerre polynomials $L_m^a (x) $ occur as the polynomial
eigenfunctions
of the generalized Laguerre operator
\begin{equation}
\label{e2.5.6}
E_2 (x, d_x) = xd_x^2 + (a+1-x) d_x
\end{equation}
where $a$ is any real number. Of course, the operator (\ref{e2.5.6}) can be
rewritten as
\begin{equation}
\label{e2.5.7}
E_2 = \ J^0_0 J^-_0 \ -\  J^0_0 \ + \ (a+1) J^-_0
\end{equation}

{\it 3. Legendre polynomials.}

The Legendre polynomials $P_{2m+k} (x)$ are the  the polynomial eigenfunctions
of the operator
\begin{equation}
\label{e2.5.8}
E_2 (x, d_x) = (1-x^2)d_x^2 -2x d_x
\end{equation}
or,
\begin{equation}
\label{e2.5.9}
E_2 = \ - J^0_0 J^0_0 \ +\ J^-_0 J^-_0\ - \ J^0_0
\end{equation}
Analogously to the Hermite polynomials there exists another way to represent
the operators
related to the Legendre polynomials. Let us notice that $k$ has a meaning of
the parity
of the polynomial  $P_{2m+k}$ and
\[
P_{2m+k} (x) = x^k p_m (x^2)
\]
Then it is easy to find that the operator having $p_m(y)$ as the eigenfunctions
\begin{equation}
\label{e2.5.10}
\bar{E}_2 (y, d_y) = 4y(1-y)d_y^2 +2[ 1+2k -(3+2k)y] d_y
\end{equation}
and, correspondently
\begin{equation}
\label{e2.5.11}
\bar{E}_2 = \ -4J^+_0 (y)J^-_0 (y)\ +\ 4J^0_0 (y)J^-_0 (y)\ -\ 2 (3+2k) J^0_0
(y) \
+\ 2 (1+2k) J^-_0 (y)
\end{equation}

{\it 4. Jacobi polynomials.}

The Jacobi polynomials appear as the polynomial eigenfunctions of Jacobi
equation
 taking in either symmetric form with the operator
\begin{equation}
\label{e2.5.12}
E_2 (x, d_x) = (1-x^2)d_x^2 + [b-a- (a+b+2)x] d_x
\end{equation}
corresponding to
\begin{equation}
\label{e2.5.13}
E_2= \ -J^0_0 J^0_0  + J^-_0 J^-_0 - (1+a+b) J^0_0 + (b-a) J^-_0,
\end{equation}
or asymmetric form  (see e.g. the book by Murphy \cite{m} or
Bateman--Erd\'{e}lyi
\cite{be})
\begin{equation}
\label{e2.5.14}
E_2 (x, d_x) = x(1-x)d_x^2 + [1+a-(a+b+2)x] d_x
\end{equation}
corresponding to
\begin{equation}
\label{e2.5.15}
E_2 = \ -J^0_0 J^0_0  + J^0_0 J^-_0 - (1+a+b) J^0_0 + (a+1) J^-_0,
\end{equation}


Under special choices of the general element $E_4 (E_6, E_8)$,
one can reproduce all known fourth-(sixth-, eighth-)order differential
equations
giving rise to infinite sequences of orthogonal polynomials (see e.g.
Littlejohn \cite{little}
and other papers in this volume).
\vskip 1.truecm

  Recently, A. Gonz\'alez-Lop\'ez, N. Kamran and P. Olver \cite{olver}
gave the complete description of second-order polynomial elements of
$U_{sl_2({\bf R})}$ in the representation (2.1)  leading to the
square-integrable
eigenfunctions of the Sturm-Liouville problem (2.21) after transformation
(2.22)-(2.23). Consequently, for second-order ordinary differential equation
(2.17) the combination of their result and Theorems 2.1, 2.2 gives a general
solution of the problem of classification of  equations possessing a finite
number
of orthogonal polynomial solutions.

\newpage
\renewcommand{\theequation}{3.{\arabic{equation}}}
\setcounter{equation}{0}
\section{ Finite-difference equations in one variable}
\subsection{General consideration}
Let us define a multiplicative finite-difference operator, or a shift operator
or the so-called Jackson symbol (see e.g. Exton \cite{e}, Gasper and Rahman
\cite{gr})
\begin{equation}
\label{e3.1}
D f(x) = {{f(x) - f(qx)} \over {(1 - q) x}}
\end{equation}
where $q \in R$ and $f(x)$ is real function $x \in R$ . The Leibnitz rule for
the
operator $D$ is
\[
D f(x) g(x)= (D f(x)) g(x)+ f(qx) Dg(x)
\]
Now one can easily introduce the finite-difference analogue of the algebra
of  the differential operators (2.1) based on the operator $D$ instead of the
continuous derivative  (Ogievetsky and Turbiner \cite{ot})
\label{e3.2}
\[ \tilde  J^+_n = x^2 D - \{ n \} x \]
\begin{equation}
\tilde  J^0_n = \  x D - \hat{n}
\end{equation}
\[ \tilde  J^-_n = \ D , \]
where $\{n\} = {{1 - q^n}\over {1 - q}}$  is so called $q$ number and
$\hat n \equiv {\{n\}\{n+1\}\over \{2n+2\}}$.
The operators (3.2) after multiplication by some factors
\[ \tilde  j^0 = {q^{-n} \over p+1} {\{2n+2\} \over \{n+1\}} \tilde J^0_n \]
\[ \tilde  j^{\pm} = q^{-n/2} \tilde  J^{\pm}_n \]
(see Ogievetsky and Turbiner \cite{ot}) form a quantum algebra  $sl_2({\bold
R})_q$
with the following commutation relations
\label{3.2}
\[ q \tilde  j^0\tilde  j^- \ - \ \tilde  j^-\tilde  j^0 \
= \ - \tilde  j^-  \]
\begin{equation}
 q^2 \tilde  j^+\tilde  j^- \ - \ \tilde  j^-\tilde  j^+ \
= \ - (q+1) \tilde  j^0
\end{equation}
\[ \tilde  j^0\tilde  j^+ \ - \ q\tilde  j^+\tilde  j^0 \ = \  \tilde  j^+  \]
The parameter $q$ does characterize the deformation of the commutators of the
classical Lie algebra  $sl_2$. If $q \rightarrow 1$, the commutation relations
(3.3)
reduce to the standard $sl_2({\bold R})$ ones. A remarkable property of
generators
(3.2) is that, {\it if $n$ is a non-negative integer, they form the
finite-dimensional
representation corresponding the finite-dimensional representation space
${\cal P}_{n+1}$ the same as of the non-deformed} $sl _2$ (see (2.1)).
For values of $q$ others than root of unity this representation is irreducible.

{\it Comment 3.1}\  The algebra (3.3) is known in literature as so-called
{\it the second Witten quantum deformation} of $sl_2$ in the classification
of C. Zachos \cite{z}).

Similarly as for differential operators one can introduce
quasi-exactly-solvable $\tilde  T_k(x,D)$ and exactly-solvable
finite-difference
operators $\tilde  E_k(x,D)$ (see definition 2.1).
\begin{LEMMA} (Turbiner \cite{tams}) {\it (i) Suppose $n > (k-1)$.  Any
quasi-exactly-solvable operator $\tilde T_k$, can be represented by a $k$th
degree polynomial of the operators (3.2). If $n \leq (k-1)$, the part
of the quasi-exactly-solvable operator $\tilde T_k$ containing
derivatives up to order $n$ can be represented by a $n$th
degree polynomial in the generators (3.2).

(ii) Conversely, any polynomial in (3.2) is quasi-exactly solvable.

(iii) Among quasi-exactly-solvable operators
there exist exactly-solvable operators $\tilde E_k \subset \tilde
T_k$.}
\end{LEMMA}

\noindent
{\it Comment 3.2} \ If we define an analogue of the universal enveloping
algebra $U_g$ for the quantum algebra $\tilde g$ as an algebra of all ordered
polynomials in generators,
then a quasi-exactly-solvable operator
$\tilde T_k$ at $k < n+1$ is simply an element of the \lq universal enveloping
algebra' $U_{sl_2({\bold R})_q}$ of the algebra $sl_2({\bold R})_q$ taken
in representation (3.2). If $k \geq n+1$, then $\tilde T_k$ is
represented as an element of $U_{sl_2({\bold R})_q}$ plus $B D^{n+1} $,
where $B$ is any linear difference operator of order not higher than
$(k-n-1)$.

Similar to $sl_2({\bold R})$ (see definition 2.3), one can introduce the
grading
of generators (3.2) of $sl_2({\bold R})_q$ (cf. (2.3)) and, hence, the grading
of monomials of the universal enveloping $U_{sl_2({\bold R})_q}$ (cf. (2.4)).
\begin{LEMMA} {\it A quasi-exactly-solvable operator
$\tilde T_k \subset U_{sl_2({\bold R})_q}$
has no terms of positive grading, iff it is an exactly-solvable operator.}
\end{LEMMA}
\begin{THEOREM} (Turbiner \cite{tams}) Let $n$ be a non-negative integer.
Take the eigenvalue problem for a linear difference operator of the $k$-th
order
in one variable
\begin{equation}
\label{e3.4}
 \tilde T_k (x,D) \varphi (x) \ = \ \varepsilon \varphi (x) ,
\end{equation}
where $\tilde T_k$ is symmetric. The problem (3.4) has $(n+1)$
linearly independent eigenfunctions in the form of a polynomial in
variable $x$ of order not higher than $n$, if and only if $T_k$ is
quasi-exactly-solvable. The problem (3.4) has an infinite sequence
of polynomial eigenfunctions, if and only if the operator is exactly-solvable
$\tilde E_k$.
\end{THEOREM}

{\it Comment 3.2} \ Saying the operator $\tilde T_k$ is symmetric, we imply
that, considering the action of this operator on a space of polynomials
of degree not higher than $n$, one can introduce a positively-defined
scalar product, and the operator $\tilde T_k$ is symmetric with respect
to it.

This theorem gives a general classification of finite-difference equations
\begin{equation}
\label{e3.5}
 \sum_{j=0}^{k} \tilde a_j (x) D^j \varphi (x) \ = \ \varepsilon \varphi(x)
\end{equation}
having polynomial solutions in $x$.  The coefficient functions must
have the form
\begin{equation}
\label{e3.6}
\tilde a_j (x) \ = \ \sum_{i=0}^{k+j} \tilde a_{j,i} x^i .
\end{equation}
In particular, this form occurs after substitution (3.2) into a general
$k$th degree polynomial element of the universal
enveloping algebra $U_{sl_2({\bold R})_q}$. It guarantees the existence of
at least a finite number of polynomial solutions. The coefficients
$\tilde a_{j,i}$ are related to the coefficients of the
$k$th degree polynomial element of the universal
enveloping algebra $U_{sl_2({\bold R})_q}$. The number of free parameters
of the
polynomial solutions is defined by the number of free parameters of a general
$k$-th order polynomial element of the universal
enveloping algebra $U_{sl_2({\bold R})_q}$.\footnote{For quantum
$sl_2({\bold R})_q$ algebra
there are no polynomial Casimir operators (see, e.g. Zachos \cite{z}). However,
in the representation (3.2) the relationship between generators analogous
to the quadratic Casimir operator
\[ q\tilde J^+_n\tilde J^-_n - \tilde J^0_n \tilde J^0_n + (\{ n+1 \}
- 2 \hat{n}) \tilde J^0_n = \hat{n} (\hat{n} - \{ n+1 \}) \]
appears. It reduces the number of independent parameters of the
second-order polynomial element of  $U_{sl_2({\bold R})_q}$. It becomes the
standard Casimir operator at $q \rightarrow 1$. } A rather
straightforward calculation leads to the following formula
\[ par (\tilde T_k) = (k+1)^2+1 \]
(for the second-order finite-difference equation $par(\tilde T^2) = 10$).
For the case of an infinite sequence of polynomial solutions the formula
(3.6) simplifies to
\begin{equation}
\label{e3.7}
\tilde a_j (x) \ = \ \sum_{i=0}^{j} \tilde a_{j,i} x^i
\end{equation}
and the number of free parameters is given by
\[ par (\tilde E_k) = {(k+1)(k+2) \over 2} + 1 \]
(for $k=2$, $par(\tilde E^2) = 7$).
The increase in the number of free parameters
compared to ordinary differential equations is due to the presence of the
deformation parameter $q$.

\subsection{Second-order finite-difference exactly-solvable equations.}

In Turbiner \cite{t3} it is implemented a description
in the present approach of the $q$-deformed Hermite, Laguerre, Legendre
and Jacobi polynomials (for definitions of these polynomials see Exton
\cite{e},
Gasper and Rahman \cite{gr}).
In order to reproduce the known $q$-deformed classical Hermite, Laguerre,
 Legendre and Jacobi polynomials
(for the latter, there exists the $q$-deformation of the asymmetric form
 (2.61) only, see e.g. Exton \cite{e}, Gasper and Rahman \cite{gr}),  one
should
modify the spectral problem (3.4):
\begin{equation}
\label{e3.8}
 \tilde T_k (x,D) \varphi (x) \ = \ \varepsilon \varphi (qx) ,
\end{equation}
by introducing the r.h.s. function the dependence on the argument
$qx$ (cf. (2.5) and (3.4)) as it follows from the book by Exton \cite{e} (see
also
Gasper and Rahman \cite{gr}). Then corresponding $q$-difference operators
having $q$-deformed classical Hermite, Laguerre,
Legendre and Jacobi polynomials as eigenfunctions (see the equations
(5.6.2), (5.5.7.1), (5.7.2.1), (5.8.3) in Exton \cite{e}, respectively)
are given by the combinations in the  generators:
\renewcommand{\theequation}{3.9.{\arabic{equation}}}
\setcounter{equation}{0}
\begin{equation}
\label{e3.9.1}
\tilde E_2 =  \tilde  J^-_0 \tilde  J^-_0 \ -\ \{ 2 \} \tilde  J^0_0 ,
\end{equation}
\begin{equation}
\label{e3.9.2}
\tilde E_2 =  \tilde  J^0_0 \tilde  J^-_0 \ -\ q^{-a-1} \tilde  J^0_0 \ +
\ (q^{-a-1}\{ a+1\}) \tilde  J^-_0 ,
\end{equation}
\begin{equation}
\label{e3.9.3}
\tilde E_2 = - q\tilde  J^0_0 \tilde  J^0_0  + \tilde  J^- \tilde  J^- + (q -\{
2 \})
\tilde  J^0_0 ,
\end{equation}
\begin{equation}
\label{e3.9.4}
\tilde E_2= - q^{a+b-1}\tilde  J^0_0 \tilde  J^0_0  + q^a \tilde  J^0_0
\tilde  J^-_0 + [q^{a+b-1}-\{a\}q^b -\{b\}] \tilde  J^0_0 + \{a\} \tilde  J^-_0
\ ,
\end{equation}
respectively.
\renewcommand{\theequation}{3.{\arabic{equation}}}
\setcounter{equation}{9}
\begin{LEMMA} {\it If the operator $\tilde T_2$ (for the definition, see
(2.15.1))
is such that
\begin{equation}
\label{e3.10}
\tilde c_{++}=0 \quad and \quad \tilde c_{+} =( {\hat n}  - \{ m \})
\tilde c_{+0} \ , \ at \ some\ m=0,1,2,\dots
\end{equation}
 then the operator $\tilde T_2$ preserves both ${\cal P}_{n+1}$ and
${\cal P}_{m+1}$, and polynomial solutions in $x$ with 8 free parameters
occur.}
\end{LEMMA}

As usual in quantum algebras, a rather outstanding situation occurs
if the \linebreak deformation parameter $q$ is equal to a primitive root of
unity.
For instance, the following statement holds.
\begin{LEMMA} {\it If a quasi-exactly-solvable operator $\tilde T_k$
preserves the space ${\cal P}_{n+1}$ and the parameter $q$ satisfies
to the equation
\begin{equation}
\label{e3.11}
q^n\ = \ 1 \ ,
\end{equation}
then the operator  $\tilde T_k$ preserves an infinite flag of polynomial spaces
$ {\cal P}_0 \subset  {\cal P}_{n+1} \subset {\cal P}_{2(n+1)} \subset \dots
\subset {\cal P}_{k(n+1)} \subset \dots $.}
\end{LEMMA}

It is worth emphasizing that, in the limit as $q$ tends to one,
Lemmas 3.1,3.2,3.3 and Theorem 3.1
coincide with Lemmas 2.1,2.2,2.3 and Theorem 2.1, respectively.
Thus the case of differential equations in one variable can be treated
as a limiting case of finite-difference ones. Evidently, one can consider
the finite-difference operators, which are a mixture of generators (3.2) with
the same value of $n$ and different $q$'s.

\newpage
\renewcommand{\theequation}{4.{\arabic{equation}}}
\setcounter{equation}{0}
\section{$2 \times 2$ matrix differential equations on the real line}
\subsection{General consideration}
This Section is devoted to a description of quasi-exactly-solvable , ${\bold
T}_k (x)$
and exactly-solvable, ${\bold E}_k (x)$,  $2 \times 2$
matrix differential operators acting on space of the two-component spinors with
polynomial components
\begin{equation}
\label{e4.1}
{\cal P}_{n+1,m+1} \ = \ \left \langle  \begin{array}{c}
x^0,x^1,\dots,x^m  \\
x^0, x^1, \dots,x^n \end{array} \right \rangle
\end{equation}
This space is a natural generalization of the space ${\cal P}_n $ (see (1.5)).
The definition of quasi- and exactly-solvable operators is also a natural
generalization of the Definition 2.1 .

Now let us introduce following two sets of $2 \times 2$ matrix differential
operators:
\label{e4.2}
\[ T^+ = x^2 \partial_x - n x + x \sigma^+\sigma^-,  \]
\begin{equation}
 T^0 = x \partial_x - {n \over 2} +{1 \over 2} \sigma^+\sigma^-\  ,
\end{equation}
\[ T^- = \partial_x \ .  \]
\[ J = -{ n \over 2} - {1 \over 2} \sigma^+\sigma^- \]
named bosonic (even) generators and
\begin{equation}
\label{e4.3}
Q \ = \ { \sigma^- \brack x\sigma^-}\ , \
{\bar Q} \ = \ { x\sigma^+ \partial_x-n\sigma^+ \brack
-\sigma^-\partial_x} .
\end{equation}
named  fermionic (odd) generators, where $\sigma^{\pm, 0}$ are Pauli matrices
in standard notation
\[
\sigma^+ =
\ \left( \begin{array}{cc}
0 & 1 \\
0 & 0
\end{array}  \right)
\ ,\ \sigma^-=
\ \left( \begin{array}{cc}
0 & 0 \\
1 & 0
\end{array}  \right)
\ ,\ \sigma^0 =
\ \left( \begin{array}{cc}
1 & 0 \\
0 & -1
\end{array}  \right) \ .
\]
It is easy to check that these generators form the algebra $osp(2,2)$:
\label{e4.4}
\[
[T^0 , T^{\pm}]= \pm T^{\pm} \quad , \quad [T^+ , T^-]= -2 T^0 \quad , \quad
 [J , T^{\alpha}]=0  \quad , \alpha = +,-,0 \]
\[   \{ Q_1,\overline{Q}_2 \}  = - T^- \quad , \quad \{ Q_2, \overline{Q}_1 \}
= T^+ \quad, \]
\[ {1 \over 2} (\{ \overline{Q}_1, Q_1\} + \{ \overline{Q}_2, Q_2 \})
= + J \quad ,
\quad
 {1 \over 2} (\{ \overline{Q}_1, Q_1\} - \{ \overline{Q}_2, Q_2 \}) =  T^0
\quad , \]
\[ \{ Q_1,Q_1\}=\{ Q_2,Q_2\}=\{ Q_1,Q_2\}=0\quad , \]
\[ \{ \overline{Q}_1,\overline{Q}_1\}=
\{ \overline{Q}_2,\overline{Q}_2\}=\{ \overline{Q}_1,\overline{Q}_2\}=0 \quad
,\]
\[  [Q_1 , T^+]=Q_2 \quad , \quad [Q_2 , T^+]=0 \quad , \quad [Q_1 ,
T^-]=0 \quad , \quad [Q_2 , T^-]=-Q_1 \ ,\]
\[  [\overline{Q}_1 , T^+]=0 \quad , \quad  [\overline{Q}_2 , T^+] = -
\overline{Q}_1 \quad , \quad
[\overline{Q}_1 ,T^-] = \overline{Q}_2 \quad , \quad [\overline{Q}_2 ,
T^-]=0 \ ,\]
\[    [Q_{1,2} , T^0]=\pm {1 \over 2} Q_{1,2} \quad , \quad
 [\overline{Q}_{1,2} , T^0]=\mp {1 \over 2} \overline{Q}_{1,2} \]
\begin{equation}
  [Q_{1,2}, J] = - {1 \over 2} Q_{1,2} \quad , \quad
 [ \overline{Q}_{1,2}, J] = {1 \over 2} \overline{Q}_{1,2}
\end{equation}
This algebra contains the algebra $sl_2(\bold R) \oplus {\bold R}$ as
sub-algebra.

\begin{LEMMA} (Turbiner \cite{tams}) {\it Consider the space
${\cal P}_{n+1,n}$.

 (i) Suppose $n > (k-1)$.  Any quasi-exactly-solvable operator
${\bold  T}_k (x)$, can be
represented by a $k$th degree polynomial of the operators (4.2), (4.3).
 If $n \leq (k-1)$, the part of the quasi-exactly-solvable operator
${\bold T}_k (x)$ containing
derivatives in $x$ up to order $n$ can be represented by an $n$th
degree polynomial in the generators (4.2), (4.3).

(ii) Conversely, any polynomial in (4.2), (4.3) is a quasi-exactly solvable
operator.

(iii) Among quasi-exactly-solvable operators
there exist exactly-solvable operators ${\bold  E}_k \subset {\bold  T}_k
(x)$.}
\end{LEMMA}

Let us introduce the grading of the bosonic generators (4.2)
\begin{equation}
\label{e4.5}
deg (T^+) = +1 \ , \ deg (J,T^0) = 0 \ , \ deg (J^-) = -1
\end{equation}
and fermionic generators (4.3)
\begin{equation}
\label{e4.6}
deg (Q_2,\overline{Q}_1) =+ {1 \over 2}\ , \
deg (Q_1,\overline{Q}_2) =- {1 \over 2}
\end{equation}
Hence the grading of monomials of the generators (31), (32) is equal to
\label{e4.7}
\[ deg [(T^+)^{n_+} (T^0)^{n_0}(J)^{\overline{n}}(T^-)^{n_-}{Q_1}^{m_1}
{Q_2}^{m_2}
{\overline{Q}_1}^{\overline{m}_1}{\overline{Q}_2}^ {\overline{m}_2} ]
 \  = \]
\begin{equation}
  (n_+ - n_-) \ - \ (m_1 - m_2 - {\overline{m}_1} + {\overline{m}_2}) / 2
\end{equation}
The $n$'s can be arbitrary  non-negative integers, while the $m$'s are
equal to either 0 or 1. The notion of grading allows us to classify the
operators ${\bold T}_k (x)$ in the Lie-algebraic sense.
\begin{LEMMA} {\it A quasi-exactly-solvable operator
${\bold T}_k (x) \subset U_{osp(2,2)}$
 has no terms of positive grading other than  monomials of
grading +1/2 containing the generator $Q_1$ or $Q_2$,
 iff it is an exactly-solvable operator.}
\end{LEMMA}

Take the eigenvalue problem
\begin{equation}
\label{e4.8}
 {\bold T}_k (x) \varphi (x)\ = \ \varepsilon \varphi (x)
\end{equation}
where
\begin{equation}
\label{e4.9}
\varphi(x) \ = \ { \varphi_1(x) \brack \varphi_2(x)} ,
\end{equation}
is a two-component spinor.

\begin{THEOREM} (Turbiner \cite{tams}) {\it Let $n$ be a non-negative integer.
Take the eigenvalue
problem (4.7), where $ {\bold T}_k(x)$ is symmetric. In general,
the problem (4.7) has $(2n+1)$ linearly independent eigenfunctions
in the form of polynomials  ${\cal P}_{n+1,n}$ ,
if and only if ${\bold T}_k (x)$ is quasi-exactly-solvable.
The problem (4.7) has an infinite sequence of polynomial eigenfunctions,
if and only if the operator is exactly-solvable.}
\end{THEOREM}

As a consequence of Theorem 4.1, the $2 \times 2$ matrix quasi-exactly-solvable
differential operator ${\bold T}_k (x)$, possessing in general $(2n+1)$
polynomial eigenfunctions of the form ${\cal P}_{n,n-1}$ can be written in the
form
\begin{equation}
\label{e4.10}
 {\bold T}_k (x)\ =\ \sum_{i=0}^{i=k} {\bold a}_{k,i} (x) d_x^i \ ,
\end{equation}
 The coefficient functions $ {\bold a}_{k,i} (x)$
are by $2 \times 2$ matrices and generically for the $k$th order
quasi-exactly-solvable operator their matrix elements are polynomials.
Suppose that $k>0$. Then the matrix elements are given
by the following expressions
\begin{equation}
\label{e4.11}
  {\bold a}_{k,i} (x) \ = \ \left( \begin{array}{cc}
A_{k,i}^{[k+i]} & B_{k,i}^{[k+i-1]} \\
C_{k,i}^{[k+i+1]} & D_{k,i}^{[k+i]}
\end{array}  \right)
\end{equation}
at $k > 0$, where the superscript in square brackets displays the order of
the corresponding polynomial.

 It is easy to calculate the number of free parameters of a
quasi-exactly-solvable operator ${\bold T}_k (x)$
\begin{equation}
\label{e4.12}
par({\bold T}_k (x))= 4 (k+1)^2
\end{equation}

For the case of exactly-solvable problems, the matrix elements (4.10)
of the coefficient functions are modified
\begin{equation}
\label{e4.13}
  {\bold a}_{k,i} (x) \ = \ \left( \begin{array}{cc}
A_{k,i}^{[i]} & B_{k,i}^{[i-1]} \\
C_{k,i}^{[i+1]} & D_{k,i}^{[i]}
\end{array}  \right)
\end{equation}
where $k > 0$. An infinite family of orthogonal polynomials as
eigenfunctions of Eqs. (4.8)-(4.9) , if they exist, will occur, if and only
if the coefficient functions have the form (4.12). The number of free
parameters of an exactly-solvable operator ${\bold E}_k (x)$ and,
correspondingly, the maximal number of free parameters of the $2 \times 2$
matrix orthogonal polynomials in one real variable, is equal to
\begin{equation}
\label{e4.14}
par({\bold E}_k (x))= 2k (k+3) + 3 \ .
\end{equation}

 Thus, the above formulas describe the coefficient functions of matrix
differential equation (4.8), which can possess polynomials in $x$ as
solutions.
\subsection{Quasi-exactly-solvable matrix Schroedinger equations (example).}
Now let us take the quasi-exactly-solvable matrix operator ${\bold T}_2 (x)$
 and try to reduce Eq. (4.8) to the Schroedinger equation
\begin{equation}
\label{e4.15}
[ -{1 \over 2} {d^2 \over dy^2} + {\bold V}(y) ] \Psi (y)\ =\ E \Psi (y)
\end{equation}
where $ {\bold V}(y)$ is a two-by-two {\it hermitian} matrix, by making a
change of variable $x \mapsto y$ and \lqq gauge" transformation
\begin{equation}
\label{e4.16}
\Psi (y) \ = \ {\bold U}(y) \varphi (x(y))
\end{equation}
where ${\bold U}$ is an arbitrary $2 \times 2$ matrix depending on
the variable $y$. In order to get some \lqq reasonable" Schroedinger
equation one should fulfill two requirements: (i) the  potential  ${\bold
V}(y)$ must be
hermitian and (ii) the eigenfunctions $\Psi(y)$ must belong to a
certain Hilbert space.

Unlike the case of quasi-exactly-solvable differential operators in one real
variable (see Gonz\'alez-Lop\'ez, Kamran and Olver\cite{olver}), this problem
has no complete solution so far.
Therefore it seems instructive to display a particular non-trivial example
of the quasi-exactly-solvable $2 \times 2$ -matrix Schroedinger operator
(Shifman and Turbiner \cite{st}).

Take the quasi-exactly-solvable operator
\[ {\bold T}_2 = - 2 T^0 T^- + 2 T^- J - i \beta T^0 Q_1 + \]
\begin{equation}
\label{e4.17}
\alpha T^0 - (2n+1) T^- - {i\beta \over 2}(3n + 1) Q_1 + {i \over 2}\alpha
\beta Q_2 - i \beta \overline{Q}_1 \ ,
\end{equation}
where $\alpha$ and $\beta$ are parameters. Upon introducing a new
variable $y=x^2$ and  after straightforward calculations
one finds the following expression for the matrix $U$ in Eq. (4.16)
\begin{equation}
\label{e4.18}
{\bold U}  = \exp ( - {\alpha y^2 \over 4} + {i \beta y^2 \over 4} \sigma_1)
\end{equation}
and for the potential ${\bold V}$ in Eq. (4.15)
\[ {\bold V} (y) = {1 \over 8} (\alpha ^2 - \beta ^2) y^2 + \sigma _2
[-(n + {1 \over 4}) \beta + {\alpha \beta \over 4} y^2 - {\alpha \over 4}
\tan {\beta y^2 \over 2}] \cos {\beta y^2 \over 2} +  \]
\begin{equation}
\label{e4.19}
\sigma _3 [-(n + {1 \over 4}) \beta + {\alpha \beta \over 4} y^2 -
{\alpha \over 4}
\cot {\beta y^2 \over 2}] \sin {\beta y^2 \over 2}
\end{equation}
It is easy to see that the potential ${\bold V}$ is hermitian;
$(2n+1)$ eigenfunctions have the form of polynomials multiplied
by the exponential factor $U$ and they are obviously normalizable.

Recent development of matrix quasi-exactly-solvable differential equations and
further
explicit examples can be found in Brihaye-Kosinski \cite{bk}.

\newpage
\renewcommand{\theequation}{5.{\arabic{equation}}}
\setcounter{equation}{0}
\section{ Ordinary differential equations with the parity operator }

Let us introduce the parity operator  $K$ in the following way
\begin{equation}
\label{e5.1}
K f(x) = f(-x)
\end{equation}
The operator $K$ has such properties
\begin{equation}
\label{e5.2}
 \{K, x \}_+ \ = 0\ ,\  \{K, d_x \}_+ \ = 0\  , \ K^2\ =\ 1.
\end{equation}
Now take the following generators
\begin{equation}
\label{e5.3}
J^-(n) = d_x + {\nu \over x}(1-K)\ ,
\end{equation}
\begin{equation}
\label{e5.4}
J_{1}^0(n) = x J^-(n) = x d_x + \nu (1-K)\ ,
\end{equation}
\begin{equation}
\label{e5.5}
J^0(n) = n -  x d_x \ ,
\end{equation}
\begin{equation}
\label{e5.6}
J^+(n) = x J^0 (n) = nx - x^2 d_x\  .
\end{equation}
These operators together with the operator $K$ form an algebra,
which was named $gl(2,{\bf R})_K$ in Turbiner \cite{tca},  with commutation
relations
\begin{equation}
\label{e5.7}
[ J^0(n), J_1^0(n)] = 0\ ,
\end{equation}
\begin{equation}
\label{e5.8}
[ J^{\pm}(n), J^0(n)] = \pm J_i^{\pm}(n)\ ,
\end{equation}
\begin{equation}
\label{e5.9}
[ J^{\pm}(n), J_1^0(n)] = \mp (1 \mp 2\nu K) J^{\pm}(n)\ ,
\end{equation}
\begin{equation}
\label{e5.10}
[ J^+(n), J^-(n)] = J_1^0(n) - (1+2\nu K) J^0(n)
\end{equation}
and also
\begin{equation}
\label{e5.11}
[K,J_1^0(n)]\ =\ [K,J^0(n)]\ =\ 0\ , \{ K, J^{\pm}(n) \}_+\ =\ 0\ .
\end{equation}
If $\nu=0$ the generators (5.3)-(5.6) and commutation relations (5.7)-(5.10)
become
those of  the algebra $gl(2,{\bf R})$ \footnote{Hereafter we will omit the
argument $n$
in the generators except for the cases where the representation (5.3)--(5.6) is
used
concretely}.

It is easy to check, that the algebra (5.7)--(5.11) possesses two Casimir
operators --
the operators commuting with all generators of the algebra $gl(2,{\bf R})_K$:
the linear one
\begin{equation}
\label{e5.12}
C_1 =  J^0+ J^0_1 + \nu K
\end{equation}
and the quadratic one
\begin{equation}
\label{e5.13}
C_2= {1 \over 2}\{ J^+, J^- \}_+ + {1 \over 4} (J^0-J_1^0)^2 -
 {\nu \over 2} (J^0-J_1^0) K + {\nu \over 2}K \ .
\end{equation}
Substituting the concrete representation (5.3)--(5.6) into (5.12)--(5.13) , one
can find that
\[
C_1 = n + \nu \ ,
\]
\begin{equation}
\label{e5.14}
C_2 =  {1 \over 4} [(n+ \nu)(n+\nu +2) - \nu^2]
\end{equation}
The representation (5.3)-(5.6) has an outstanding property: for generic $\nu$,
when $n$ is a non-negative integer number, there appears a
finite-dimensional irreducible representation ${\cal P}_{n+1}$ of dimension
$(n+1)$.
So the representation space remains the same for any value of parameter $\nu$.

Like it has been done before one can introduce quasi-exactly-  and
exactly-solvable
mixed operators,  containing the differential operator and the parity operator
$K$ :
$T_k ( x,d_x,K)$ and $E_k( x,d_x,K)$, respectively.
It is evident, that those operators are linear in  $K$.

\begin{LEMMA}
(i) Suppose $n > (k-1)$.  Any quasi-exactly-solvable
operator $T_k$, can be
represented by a $k$-th degree polynomial of the operators (5.3) - (5.6)
If $n \leq (k-1)$, the part of the quasi-exactly-solvable operator $T_k$
containing derivatives up to order $n$ can be represented by an $n$th
degree polynomial in the generators (5.3)-(5.6).

(ii) Conversely, any polynomial in (5.3)-(5.6) is quasi-exactly solvable.

(iii) Among quasi-exactly-solvable operators
there exist exactly-solvable operators $E_k \subset T_k$.
\end{LEMMA}

Similarly to the case of  $sl_2({\bold R})$ one can introduce the grading
of generators (5.3)-(5.6)
\begin{equation}
\label{e5.15}
deg (J^+(n)) = +1 \ , \ deg (J^0(n),J^0_1(n),K) = 0 \ , \ deg (J^-(n)) = -1 ,
\end{equation}
and
\begin{equation}
\label{e5.16}
deg [(J^+(n))^{k_+} (J^0(n))^{k_0}(J^0_1(n))^{k_{0,1}} (K)^{k} (J^-(n))^{k_-}]
\
= \ k_+ - k_- .
\end{equation}
(cf. (2.3)--(2.4)).
The grading allows us to classify the operators $T_k$ in the
algebraic sense.
\begin{LEMMA} {\it A quasi-exactly-solvable operator $T_k \subset
U_{gl_2({\bold R})_K}$ has no terms of positive grading, if and only if
it is an exactly-solvable operator.}
\end{LEMMA}

\begin{THEOREM}  Let $n$ be a non-negative integer. Take the eigenvalue
problem for a linear differential
operator of the $k$th order in one variable
\begin{equation}
\label{e5.17}
 T_k(x,d_x,K) \varphi \ = \ \varepsilon \varphi \ ,
\end{equation}
where $T_k$ is symmetric. The problem (5.17) has $(n+1)$ linearly independent
eigenfunctions in the form of a polynomial in variable $x$ of order
not higher than $n$, if and only if $T_k$ is quasi-exactly-solvable.
The problem (5.17) has an infinite sequence of polynomial eigenfunctions,
if and only if the operator is exactly-solvable.
If  $T_k$ has no terms of odd grading, $(k_+ - k_-)$ is odd number (see
(5.16)),
$T_k$ commutes with $K$ and eigenfunctions in (5.17) have definite parity with
respect to
$x \rightarrow -x$.
\end{THEOREM}

Following the Lemma 5.1 a general second-order quasi-exactly-solvable
differential
operator is defined by a quadratic polynomial in generators of $gl_2({\bold
R})_K$.
Provided that the conditions (5.12)--(5.14) are taken into account
\footnote{ It leads to a disappearance  of the terms containing, for instance,
$J_1^0(n)$
and $J^0(n)  J^0(n)$} we arrive at
\newpage
\[
T_2 =  c_{++} J^+(n) J^+(n)  + c_{+0} J^+(n)  J^0(n) + c_{+-} J^+(n)  J^-(n)
+
\]
\renewcommand{\theequation}{5.18.{\arabic{equation}}}
\setcounter{equation}{0}
\begin{equation}
\label{e5.18.1}
 c_{0-} J^0(n)  J^-(n)  + c_{--} J^-(n)  J^-(n)  + c_+ J^+(n)  + c_0 J^0(n) +
c_- J^-(n)  + c ,
\end{equation}
(cf. (2.15.1), where $c_{\alpha \beta}= C^0_{\alpha \beta}+C^K_{\alpha \beta},
c_{\alpha}=C^0_{\alpha}+C^K_{\alpha}, c=C^0+C^K $ and all $C$'s are real
numbers.
The number of free parameters is $par (T_2) = 18$. Non-existence in $T_2$
of the terms of odd grading leads to the conditions
\[
c_{+0}\ =\ c_{0-}\ =\ 0
\]
and
\[
c_{+}\ =\ c_{-} \ =\ 0
\]
and, finally, a general operator having eigenfunctions of a definite parity is
\begin{equation}
\label{e5.18.2}
T_2^{(e,o)} =  c_{++} J^+(n) J^+(n)  +  c_{+-} J^+(n)  J^-(n)  +
 c_{--} J^-(n)  J^-(n)  + c_0 J^0(n) + c ,
\end{equation}
and the number of free parameters is $par (T_2^{(e.o)}) = 10$.

The condition of Lemma 5.2 requires that
\[
c_{++}\  =\ c_{+0} \ = \ c_+\  =\ 0 \ ,
\]
and then  the operator $T_2$ becomes exactly-solvable
\begin{equation}
\label{e5.18.3}
E_2 =   c_{+-} J^+(n)  J^-(n)  +  c_{0-} J^0(n)  J^-(n)  +
c_{--} J^-(n)  J^-(n)  + c_0 J^0(n)  + c_- J^-(n)  + c ,
\end{equation}
(cf. (2.15.2) and the number of free parameters is
reduced to $par (E_2) = 12$.

\renewcommand{\theequation}{5.{\arabic{equation}}}
\setcounter{equation}{18}

\end{document}